# Analysis of ray stability and caustic formation in a layered moving fluid medium[⊥]

David R. Bergman[*]

Morristown NJ

*Abstract*

Caustic formation occurs within a ray skeleton as optical or acoustic fields propagate in a medium with variable refractive properties and are unphysical, their presence being an artifact of the ray approximation of the field, and methods of correcting the field near a caustic are well known. Differential geometry provides a novel approach to calculating acoustic intensity, assessing ray stability and locating caustics in acoustic ray traces when the properties of medium are completely arbitrary by identifying points on the caustic with conjugate points along various rays. The method of geodesic deviation is applied to the problem of determining ray stability and locating caustics in 2-dimensional acoustic ray traces in a layered moving medium. Specifically, a general treatment of caustic formation in sound ducts and in piecewise continuous media is presented and applied to various idealized and realistic scenarios.

---

# I INTRODUCTION

The application of differential geometry to the problem of acoustic ray theory offers a unique way to trace caustics which has distinct advantages over traditional methods [1]. The equations for the acoustic rays in a layered moving fluid medium can be solved in terms of depth integrals, the final result giving horizontal range and travel time as a function of depth and the initial conditions at the source. Caustics are located by explicitly varying the range with respect to an initial ray parameter, usually the launch angle, and searching for critical points of this variation [2]. In paraxial procedures this deviation is determined by a second order linear equation for the Jacobi field along the ray [1], [3]. The Jacobi field of the ray system is identical to the geodesic deviation vector associated with the differential geometric structure which arises from application of the method of characteristics to the equations of hydrodynamics [4]. This relation connects the conjugate point theorems of differential geometry [5] to the phenomena of caustic formation.

Caustic formation is a divergent artifact of ray theory indicating a breakdown of the existence of a unique solution of the ray equation between two points leading to an infinite value for the field amplitude in the limit of geometric optics or acoustics. While it is feasible to describe the acoustic field solely in terms of normal modes or in some cases exact solutions to the field equations, thus bypassing the occurrence of caustics, rays have many distinct advantages over fields and modes in its relative simplicity, conceptual tangibility and computational information. Furthermore, modern treatments of caustics provide accurate correction terms to the field near a caustic [6], [7], hence their precise location in space and time is needed.

By analyzing the Jacobi equation associated with the rays one gets immediate information about the focusing properties of a given medium including ray stability (i.e. whether ray converge or diverge from their neighboring rays with similar initial conditions) and caustic location. Caustic formation can occur as a result of: (a) the smooth behavior of the local sound-sped profile (SSP) or wind profile, (b) reflection from a boundary, (c) initial conditions or, (d) the presence of jump discontinuities in the derivatives of the SSP and wind profile. Although the later case (d) may be ruled out on physical grounds its effect is of interest since piecewise continuous environmental parameters are sometimes used to model underwater and atmospheric acoustical systems. In such cases discontinuities in the sound speed and velocity gradients leads to the presence of a Dirac delta function in the Jacobi equation leading to special boundary conditions for matching the Jacobi fields of different segments at the boundary between two regions. In this article a complete treatment of ray stability and the formation of caustics in the acoustic field propagating in a 2 dimensional layered moving fluid background is presented.

**II THEORY**

**IIa RAY THEORY**

The ray equations in a moving layered fluid medium are well known, having been presented in its general form, derived from the theory of characteristics [3], [8], [9]. The Cartesian coordinates and the travel time are expressed here in the form of a space – time trajectory and parameterized by an arbitrary parameter, $\lambda$ [1]. Consider a medium with sound speed $c(z)$ and one dimensional fluid velocity of the form $w_x(z) \equiv w(z)$, where $z$ is depth or height. Rays that are initially fired in the $x - z$ plane do not turn out of their initial osculating plane, *i.e.* are torsion free, and constitute an effective two dimensional system. The range, depth and travel time are given by a first order system,

$$\dot{x} = \kappa_0(\alpha + c^{-2}w(1-w\alpha)) , \tag{1}$$

$$\dot{z} = \pm\kappa_0 c^{-1}\sqrt{(1-w\alpha)^2 - \alpha^2 c^2} , \tag{2}$$

$$\dot{t} = \kappa_0 c^{-2}(1-w\alpha) , \tag{3}$$

in which dot denotes differentiation with respect to $\lambda$, $\alpha = \cos\theta_0/(c_0 + \upsilon_0\cos\theta_0)$ is the ray parameter[2] and the integration constant $\kappa_0$ is chosen such that $(dt/d\lambda)_0 = 1$. Presented in this form the rays or bicharachteristics are null geodesics of a pseudo-Riemannian manifold[3]. This connection has been pointed out independently by White [9] and Unruh [10] and served as a primary inspiration for the work in this article.

---

[1] Strictly speaking this ray parameter is an affine parameter. This designation is required to ensure that the rays are in fact geodesics.

## IIb JACOBI EQUATION, GEODESIC DEVIATION

The spreading or stability equation for the system of Eqs. (1) – (3) is

$$\frac{d^2Y}{d\lambda^2} + KY = 0 \tag{5}$$

in which,

$$K = \kappa_0^2 (\alpha\, c^{-2} \upsilon'' (1-\alpha\upsilon) - \alpha^2 c^{-2} (\upsilon')^2 - \alpha\, c^{-3} c' \upsilon' (1-\alpha\upsilon) + \alpha^2 c^{-1} c'') \ . \tag{6}$$

---

[2] $\theta_0$ is the initial angle between the wavefront normal and the $x$ – axis.
[3] This is similar to the situation in general relativity where these special curves describe the trajectory of photons in a curved space – time.

The quantity $K$ measures the stability of the ray system along a given ray, labeled by $\alpha$. The individual terms in Eq. (6) directly affect the convergence of neighboring rays in a predictable way. Equation (5) has a general solution,

$$Y = \dot{z} c k_1 \int c^{-2} \dot{z}^{-3} dz + \dot{z} c k_2 \ , \tag{8}$$

with integration constants $k_1$ and $k_2$ set to match the initial conditions. To model sound from a point source the appropriate initial conditions are $Y_0 = 0$ and $\dot{Y}_0 = c_0 \delta\theta$, where $c_0 \equiv c(z_0)$ and $\delta\theta$ is an arbitrary initial deviation in ray launch angle. Geometrically the deviation, $Y$, is tangent to the wavefront and gives a local measure of the deformation of the wavefront. Equation (5) is the Jacobi equation of the ray system defined by Eqs (1)-(3). From the interpretation of rays as null geodesics it follows that $K$ is a direct measure of the curvature of the manifold defined by the method of characteristics and Eq (5) is the equation of geodesic deviation, [5], [11]. For two dimensional acoustics problems $Y$ can be used to calculate the intensity of the acoustic field from the conservation law $I(Y\hat{n}\cdot\hat{t}) = \text{constant}$, in which $\hat{n}\cdot\hat{t}$ is the projection of the ray tangent, $\hat{t}$, onto the wavefront normal, $\hat{n}$ [12], [13].

## IIc CAUSTICS AND CONJUGATE POINTS

A caustic is the locus of points determined by solutions of the equation $Y = 0$. Solving this equation gives $\lambda_c(\alpha, z_0)$, the value of the ray parameter (affine parameter) at the caustic location in terms of the ray's initial conditions. Inserting this value into the ray coordinates then gives the caustic as a curve in space, $x_c(\alpha, z_0)$, $z_c(\alpha, z_0)$, parameterized by the initial conditions of the ray. Based on the comments of the preceding section, at these points the intensity becomes infinite. This divergent artifact signals a breakdown in the validity of ray theory and is corrected for by field expansions near the caustic.

From the point of view of differential geometry the vanishing of an otherwise nontrivial deviation vector, *Y*, at two points along a geodesic indicates a breakdown in the uniqueness of solutions to the geodesic equation[4]. While the general solution given in Eq (8) along with the appropriate initial conditions can be used to generate $Y(\lambda)$ and its behavior studied directly, much useful information can be obtained by studying the Eq (5).

There are three limiting cases for which a solution to Eq (5) can be found in terms of ordinary functions along with a simple geometric interpretation. These cases, labeled I, II and III are described by *K* = constant > 0, *K* = 0 and *K* constant < 0 respectively. Defining $\omega = \sqrt{K}$, the solution to Eq (5) in each case is

$$Y = \begin{cases} A\sin(\omega\lambda + \varphi) & \omega^2 > 0 \\ A\lambda + B & \omega^2 = 0 \\ A\sinh(\omega\lambda + \varphi) & \omega^2 < 0 \end{cases}, \tag{9}$$

in which $A$, $B$ and $\varphi$ are integration constants. An example of a manifold corresponding to each case is: sphere ($K$ = constant > 0), plane ($K$ = 0) and pseudo-sphere ($K$ constant < 0 ). In case I the deviation vector has periodic zeros signaling the onset of conjugate points (on an actual sphere or globe this would correspond to the North and South poles which are passed periodically as one travels along any longitude). A higher value of $K$ produces a higher frequency of conjugate point formation. The period between consecutive conjugate points, measured in units of $\lambda$, is given by $\Lambda_c = \pi K^{-1/2}$. In the other two cases conjugate points will not form.

---

[4] For a more complete presentation of the conjugate point theorems, geodesic deviation and Jacobi theory the reader is referred to Reference [5].

When the curvature depends on position we can say the following: if $K(z) \leq 0 \;\; \forall z$ conjugate points will never form along the ray, if $K(z) > 0 \;\; \forall z$ conjugate points will form along the ray as long as it obeys the conditions of completeness[5]. Incompleteness can occur when an absorbing boundary is present in the environment, in which case the ray may simply terminate before the caustic gets a chance to form, or because the solution to Eq (1-3) is not well defined for all $\lambda$. A simple example of the latter case occurs for the SSP $c(z) = c_0(1 + \varepsilon z)$. The exact solution for the rays is well known in this case [14] and can be used to verify that the final value of $\lambda$ is finite for any ray launched at an angle $-\pi/2 \leq \theta_0 < \pi/2$ from a source placed at a finite value of $z_0$ and landing at $z_f = -\varepsilon^{-1}$. Of particular importance are cases when, due to the concavity of the environmental parameters, rays become trapped between 2 vertical turning points. These trapped rays will, ideally, propagate forever in the horizontal direction as they oscillate in the vertical direction. If $K > 0$ everywhere along the ray the conjugate point theorem states that the period in λ between consecutive conjugate points obeys the relation $\pi K_{Max}^{-1/2} < \lambda_c < \pi K_{min}^{-1/2}$, where $K_{min} < K < K_{Max}$. It is precisely this scenario that we consider in the next section.

---

[5] A geodesic is complete if its points exist for all $\lambda \in (-\infty, \infty)$.

## III APPLICATION TO SMOOTH *c(z)* and *w(z)* SOUND DUCTS

The formalism of section **II** is applied to an environment with a smoothly varying[6] SSP and horizontal current (or wind), each a function of depth, *z*. To better understand the effects of the environment on the acoustic field consider Eq. (5) in detail. Each term in Eq. (6) has a distinct effect on the field described as follows[7].

The last term, $c''$, governs the focusing properties of the medium caused by an inhomogeneous sound speed. A ray propagating in a region with $c'' > 0$ will eventually encounter conjugate points, while rays propagating in regions where $c'' < 0$ will diverge from one another. The second term, proportional to $\alpha^2 c^{-2}(w')^2$, is always negative causing ray divergence. The first term, $\alpha w''$, will cause focusing of acoustic rays when $\alpha$ and $w''$ are the same sign. When both *c* and *w* depend on depth the term $-\alpha c^{-1} c' w' \dot{t}$ in *K* couples the sound-speed gradient to the fluid-velocity gradient. Consider a simple situation in which a waveguide is created by a sound-speed profile with $c'' > 0$ everywhere and $c' > 0$ ($c' < 0$) above (below) the waveguide axis. Furthermore let the fluid velocity satisfy $w > 0$ and $w' > 0$ everywhere in the waveguide. For acoustic rays with $\alpha > 0$, there is a separation of neighboring rays above the waveguide axis and an enhanced convergence of rays below the waveguide axis. When the background fluid motion is weak and slowly varying, the leading-order terms in Eq. (6) are $\alpha c^{-2} w'' + \alpha^2 c^{-1} c''$, indicating that the dominant effects are due to the concavity of the environmental parameters.

---

[6] Here we mean *c*, *w*, and all derivatives are continuous.
[7] This description appears in Ref [1] and is added here for completeness. In the following description an overall factor of $\kappa_0^2$ is ignored.

The sound-speed and wind gradients affect the bending and twisting of the acoustic rays. In general rays may be unbound or bound depending on how the environmental parameters vary with position. A common example of this is given by the Munk profile which creates a natural sound duct or channel [15]. From the comments of the previous paragraph it is clear that ducted regions may be created in the atmosphere or ocean by the concavity of either the wind or sound-speed.

Consider an environment with smooth $c(z)$ and $w(z)$ such that at some depth, $z_0$: $c_0' = w_0' = 0$, $c_0'' > 0$ and $w_0'' \neq 0$. These conditions define a sound duct (wave guide) at $z = z_0$, near which $c \approx c_0 + c_0''(z - z_0)^2/2$ and $w \approx w_0 + w_0''(z - z_0)^2/2$. Furthermore, since $c_0' = w_0' = 0$, a point source placed on the sound duct axis, $z = z_0$, will launch two special rays described by initial conditions $p = \pm 1$, which travel along the sound channel axis in the $\pm x$ directions. These two rays are described exactly by: $t = \lambda$, $z = z_0$ and $x - x_0 = (w_0 \pm c_0)t$. The exact solution to Eqs (1) - (3) for quadratic $c$ and $w$ and arbitrary initial conditions involve incomplete elliptic integrals. An exact solution to equation 5 may be developed along the sound channel axis thus allowing a paraxial description of the near axis rays in terms of ordinary functions, Eq. (9). Evaluating the sectional curvature, $K$, along the rays on the sound channel axis gives the following constant

$$\omega^2 \equiv K = c_0(c_0'' \pm w_0'') \ , \tag{10}$$

for right/left axis rays. Figure 1 illustrates sample ray fans about the sound channel axis (-36° to +36°, in 6° increments) for $c \sim 1 + z^2$ and $w \sim bz^2$. Column A/B illustrates rays fired to the left/right (against/with the wind). The rows (a) to (e) correspond to increasing values of the parameter $b$ = (0, 0.2, 0.4, 0.8 and 1) respectively. In all 5 cases the flow is subsonic. The rightward ray fan has enhanced convergence due to the presence of the wind while the leftward ray fan shows the opposite effect. As the concavity of $w$ is increased the ray fans suffer more severe aberration until at $b = 1$ the caustics in the left ray fan are completely destroyed and rays near the axis begin to diverge linearly. At $b = 1$ the convergence in the left ray fan is completely destroyed. A detailed application of equation 10 allows one to predict exactly where the tips of the caustic curves on the duct axis occur when $\omega^2 > 0$, $x_c = n\pi 2^{-1/2}(1 \pm b)^{-1/2}$. Figure 2 compares the convergence and divergence zones for $b = 0$ and $b = 1$, for other values of $b$ the zone boundaries depend sensitively on the individual ray coordinates.

A striking feature of this behavior is that it is completely determined by the concavity of the wind profile and not at all by the local Mach number (in fact for the above example the source is placed in a stationary region, $w$ = 0m/s, and the concavity in $w$ completely destroys the caustic structure for $p = -1$ and $b = 1$). Hence, in general even very weak but rapidly changing winds can wreak havoc on the stability of rays in a sound duct.

To connect these analyses to a more realistic example consider a scenario in the ocean in which a sound duct is described by the Munk profile along with an oscillating current:

$c(z) = c_0(1 + \varepsilon(e^{-D} - 1 + D))$ and $w(z) = w_0 e^{-az} \sin(\beta(z-d) + \varphi)$, in which

$D = 2(z-d)/d$, $\beta \equiv 2\pi\lambda^{-1}$ and $\tan\varphi = \beta/\alpha$, where $\lambda$ is a characteristic wavelength of the current variations (not the affine parameter)[8]. The current, $w$, is fixed so the conditions described in the previous example hold. From these profiles $c_0'' = 4d^{-2}c_0\varepsilon$ and $w_0'' = -w_0\beta\sqrt{1+(\beta/\alpha)^2}$. The SSP is held fixed with $d = 1500$m, $c_0 = 1500$m/s and $\varepsilon = 0.01$ (an exaggerated Munk profile). From Eq. 10 the free parameters of the wind profile may be fixed in such a way that $K = 0$ in one direction of the waveguide. This has been done here by choosing $\alpha = 0.001$m$^{-1}$ and calculating $w_0$ for some choice of $\lambda$. Figure 3 shows wind profiles for (b) $\lambda = d$ ($w_0 = 6.632$m/s), (c) $\lambda = 2d$ ($w_0 = 24.6$m/s) and (d) $\lambda = d/2$ ($w_0 = 1.69$m/s) along with the SSP (a). Figure 4 shows a sample ray trace of the near axis rays for the choice in Fig. (3b) and point source placed at $z = 1500$m. Ray fans for a point source on the waveguide axis launched into and against the current show immense disparity. Clearly these currents are negligible compared to the local sound speed of 1500m/s and produce minor correction to travel times along rays near the wave guide axis. In spite of this the combined focusing properties of $c$ and $w$ on the axis cause serious changes in the full acoustic field indicating that these minor currents cannot be ignored in intensity calculations. Figure (5) shows the exact same ray traces as in Fig. (4) with the fluid speed increased by a factor of 2. This increase in amplitude causes $K < 0$ along the waveguide. Comparing Fig. (4c) to Fig. (5c) one can clearly see the initial exponential rate of divergence for rays near the axis in Fig. (5) as compared to the linear divergence illustrated in Fig. (4).

---

[8] This current profile is chosen for illustrative purposes since it is easy to implement and demonstrates the profound effect that a small current can have on ray stability. Similar profiles describe fluid flow in Ekman layers, for example see Ref [16]. Stratified currents of this form have been observed in the Indian ocean during monsoon season, see Ref [17].

## IV PIECEWISE CONTINUOUS PROFILES, GENERAL TREATMENT

Consider a piecewise model of an ocean or atmospheric environment in which the medium is divided in depth into a finite number of regions, see Fig. (6). Ray segments are indexed in order of increasing ray parameter without reference to the corresponding region. The Jacobi field is a function of affine parameter evaluated along the ray path. Each ray segment in space corresponds to an increment of affine parameter. The SSP and wind profile take the form

$$c(z) = c_1(z)\Theta(L_1 - z) + \sum_{i=1}^{N-2} c_{i+1}(z)\Theta(L_{i+1} - z)\Theta(z - L_i) + c_N(z)\Theta(z - L_{N-1}) \quad , \quad (11)$$

$$w(z) = w_1(z)\Theta(D_1 - z) + \sum_{i=1}^{N-2} w_{i+1}(z)\Theta(D_{i+1} - z)\Theta(z - D_i) + w_N(z)\Theta(z - D_{N-1}) \quad (12)$$

with $c_i(L_i) = c_{i+1}(L_i)$ and $w_i(D_i) = w_{i+1}(D_i)$. The first and second derivatives of $c$ and $w$ are

$$c' = c'_1\Theta(L_1 - z) + \sum_{i=1}^{N-2} c'_{i+1}\Theta(L_{i+1} - z)\Theta(z - L_i) + c'_N\Theta(z - L_{N-1})$$

$$w' = w'_1\Theta(D_1 - z) + \sum_{i=1}^{N-2} w'_{i+1}\Theta(D_{i+1} - z)\Theta(z - D_i) + w'_N\Theta(z - D_{N-1})$$

$$c'' = c''_1\Theta(L_1 - z) + \sum_{i=1}^{N-2} c''_{i+1}\Theta(L_{i+1} - z)\Theta(z - L_i) + c''_N\Theta(z - L_{N-1}) + \sum_{i=1}^{N-1} \Delta c'_i \delta(z - L_i)$$

$$w'' = w_1''\Theta(D_1 - z) + \sum_{i=1}^{N-2} w_{i+1}''\Theta(D_{i+1} - z)\Theta(z - D_i) + w_N''\Theta(z - D_{N-1}) + \sum_{i=1}^{N-1} \Delta w_i'\delta(z - D_i)$$

respectively, in which $\Delta f_i = f_{i+1} - f_i$. Equation (5) is solved in each region with the curvature term determined by the Heaviside functions $\Theta$ appearing in the above expressions. The boundary conditions at each occurrence of an interface determine the constants of integration for the Jacobi field in terms of the initial conditions. The Jacobi field is continuous and the discontinuity in $\dot{Y}$ is determined by integrating Eq. (5) along the ray as it passes across a boundary.

Two distinct ways of analyzing this problem arise that require a slightly different treatment. In the first case it is assumed that Eq. (5) may be solved to give $Y(\lambda)$ and boundary conditions are applied to points along the $\lambda$ axis as described in Fig (2). In the second case it is assumed that the solution is in the form $Y(z)$ given by Eq. (8). Boundary conditions are applied at a depth where an interface different layers occurs.

### IVa  CASE 1, BOUNDARY CONDITIONS IN $\lambda$

Passing from one region to another in space corresponds to boundary point along the $\lambda$ axis, see Fig. (7). The first boundary condition is $Y_{i+1}(\lambda_B) = Y_i(\lambda_B)$. Integrating Eq. (5) across a boundary leads to the following condition on the discontinuity in $\dot{Y}$,

$$\dot{Y}_{i+1}(\lambda_B) - \dot{Y}_i(\lambda_B) + QY_i(\lambda_B) = 0 , \tag{13}$$

$$Q = \frac{\kappa_0^2}{|\dot{z}_B|c_B^2}\left(\alpha\Delta\upsilon'(1 - \alpha\upsilon_B) + \alpha^2 c_B \Delta c'\right), \tag{14}$$

where the subscript B means evaluated at the boundary and $\Delta \upsilon'$, $\Delta c'$ are the discontinuities in the fluid velocity and sound speed respectively at the boundary. Since Y is a solution to a second order equation the values of $Y_i(\lambda_B)$ and $\dot{Y}_i(\lambda_B)$ completely determine the solution along the *i*-th segment of the ray. Once the rays are matched up the parameter value, time of flight and range for each segment are found using solutions to the ray equation.

## IVb  CASE 2, BOUNDARY CONDITIONS IN *z*

We first express the solution $Y(z) = k_1 f(z) + k_2 g(z)$, where $g(z) = c\dot{z}$ and $f(z) = c\dot{z} \int c^{-2} \dot{z}^{-3} dz$. Layers are indexed according to height along the z axis from bottom to top, the solution in each region being $Y_i(z) = k_{1,i} f_i(z) + k_{2,i} g_i(z)$. A given ray will pass through one region more than once, perhaps an infinite number of times. The bare solution along each of these segments is the same but the constants of integration will be different. Applying boundary conditions at a layer interface, with $\dot{Y} = \dot{z}Y'$, gives the coefficients if the $Y_{i+1}$ in terms of those of $Y_i$.

$$\begin{pmatrix} k_{1,i+1} \\ k_{2,i+1} \end{pmatrix} = \begin{pmatrix} D^{-1}(f_i g'_{i+1} - f'_i g_i) + q\, f_i g_i & D^{-1} g_i \Delta g'_i + q\, g_i^2 \\ -D^{-1} f_i \Delta f'_i - q\, f_i^2 & D^{-1}(f_i g'_i - f'_{i+1} g_i) - q\, f_i g_i \end{pmatrix} \begin{pmatrix} k_{1,i} \\ k_{2,i} \end{pmatrix} \quad (15)$$

in which $D = f_i g'_{i+1} - f'_{i+1} g_i$, $q = D^{-1} \dot{z}_B^{-1} Q$ and the identities $f_{i+1} = f_i$ and $g_{i+1} = g_i$ evaluated at the boundary have been used to simplify as many terms as possible. The coefficients of *Y* along any ray segment are expressed in terms of the initial conditions by repeated application of Eq. (15).

## IVc EXAMPLE

Piecewise linear profiles are particularly easy to deal with since the rays and deviation vector may be expressed in closed form in terms of ordinary functions. The effect of the jump discontinuities that these profiles produce in *K* on the acoustic field described in the last sections is illustrated here for two cases where sound from a point source placed in a homogeneous stationary medium, $c$ = constant, $w = 0$ for $z > 0$ is incident on an inhomogeneous half space: case A, $c = c_0(1 + z/L)$, $w = 0$ and case B, $c$ = constant, $w = w_0 z/L$, for $z \leq 0$. The rays in each inhomogeneous half are well known. The deviation vector for $z < 0$ is $Y_1 = A_1 + B_1 \lambda$ and for $z \geq 0$ $Y_2 = A_2 + B_2 \lambda$ for case A, $Y_2 = A_2 \cosh(\omega \lambda) + B_2 \sinh(\omega \lambda)$, in which $\omega = \kappa_0 \alpha\, w' c^{-1}$, for case B. Figure 9 illustrates a ray fan and caustics for case A with $L = 1$ and $c_0 = 1$. The caustic curve in the homogeneous space is given by

$$z_c = 2(p^{-2} - 1) - z_0, \qquad x_c = 4\sqrt{p^{-2} - 1} \tag{16}$$

or simply

$$z_c = \frac{x_c^2}{8} - z_0 \tag{17}$$

for the caustic curve in the upper half space [18]. The caustic curve in the lower portion of space is determined by the roots of the following equation for $p$

$$(a^2 + b^2)p^6 - b(2b+1)p^4 + (2b+1)p^2 - 1 = 0 \tag{18}$$

in which $b \equiv 1 + z_0$ and $a \equiv z_0^2/x_c$. The caustic as a parameterized curve $(x_c(p), z_c(p))$ for fixed $z_0$ is

$$x_c = \frac{p^3}{\sqrt{1-p^2}} \frac{z_0^2}{(z_0+1)p^2 - 1}, \qquad (z_c - 1)^2 = p^{-2}\left(1 - \frac{(1-p^2)^3}{((z_0+1)p^2 - 1)^2}\right). \tag{19}$$

A cusp forms in the lower half space which can be located by finding values of initial launch angles for which the tangent vector of the caustic curve vanishes. Differentiating Eq. (19) gives the following for the caustic tangent

$$\frac{dx_c}{dp} = z_0^2 \frac{p^2(p^2(z_0+3) - 3)}{(1-p^2)^{3/2}(p^2(z_0+1) - 1)^2}, \tag{20a}$$

$$\frac{dz_c}{dp} = z_0^2 \frac{p(p^2(z_0+3) - 3)}{((z_0+1)p^2 - 1)^{5/2}(p^4 + p^2(z_0^2 + 2z_0 - 2) + 1 - 2z_0)}. \tag{20b}$$

Both vanish when $p = \sqrt{3/(z_0 + 3)}$. This cusp will always exist in the inhomogeneous space as long as $z_0 > 0$. Ray traces and caustics are shown for a source placed at 10m in

Figure 9. The caustics, appearing on top of the ray trace, were derived from the solutions presented here. One can see the development of the cusp in the lower half space. The tail of the caustic approaches zero as $x_c \to \infty$. In the limiting case where the source in placed on the *x* axis the cusp moves right up to the source and the tail merges with the *x* axis.

Both cases, A and B, described above produce a cusp in the inhomogeneous half space. In case B this only occurs for rays launched in the same direction as the wind while those launched in the opposite direction eventually turn in the direction of the wind and do not return to the homogeneous region. Applying that same procedure on $Y(\lambda)$ for case B leads to a very lengthy expression for the caustic curve which is omitted here in the interest of brevity.

**V DISCUSSION AND CONCLUSION**

In this article a new method of determining caustic formation in layered media is presented. The method presented here generalizes to three dimensional ray tracing and four dimensional space-time ray tracing with SSP and wind depending on all three Cartesian coordinates and time [1]. A significant feature of the application to layered media is the existence of a solution to the deviation equation (Jacobi equation), Eq. (8). This solution may be included with the standard range and travel time integrals used in oceanic and atmospheric acoustics towards constructing a full ray theoretic version of the acoustic field. The application discussed here is its use in determining caustics as parameterized curves in space and judging ray stability which has distinct advantages to other approaches in its conceptual tangibility and computational use. The author has implemented Eq. (5) in a numeric dynamic ray trace procedure with the result that computation time was reduced compared to the technique of numerically differentiating the ray paths. From the identification of the stability parameter $K$ with the Gaussian curvature the conjugate point theorems provide immediate computational value and conceptual insights into the behavior of caustics.

**ACKNOWLEDGEMENTS**

The author thanks the Office of Naval Research and the American Society for Engineering Education for hosting a summer faculty fellowship at the Naval Research Laboratory (NRL) in Washington DC for the summer 2004, during which time most of this work were completed.

**FIGURE CAPTIONS**

1. Sample ray fans for $c(z)$ and $w(z)$ quadratic profiles described in the text for a point source placed at $z = 0$. Column A/B corresponds to rays launched against/with the flow of the fluid while the rows (a) – (e) correspond to increasing values of $b = (0, 0.2, 0.4, 0.8, 1.0)$.

2. Maps of the convergence/divergence zones and curves of zero curvature determined by the section curvature, $K$, for (a) case (a) and (b) case (e) of figure 1. The bold lines are curves of zero curvature. These zones depend on the source placement.

3. (a) Canonical Munk profile, $c(z) = c_0(1 + \varepsilon(e^{-D} - 1 + D))$, $D = 2(z - d)/d$ with $c_0 = 1500\text{m/s}$, $d = 1500\text{m}$ and $\varepsilon = 0.01$. (b) through (d) Oscillating current $w(z) = w_0 e^{-\alpha z} \sin(\beta(z-d) + \varphi)$, $\alpha = 0.001\text{m}^{-1}$ and (b) $\beta = 2\pi/d$ and $w_0 = 6.632$m/s, (c) $\beta = \pi/d$ and $w_0 = 1.69$m/s, (d) $\beta = 4\pi/d$ and $w_0 = 24.6$m/s.

4. Ray fans from a point source placed in an ocean environment, described by SSP and current form fig 3 (a) and (b), at $z = 1500$m, wave guide axis. (a) against the current, (c) with the current, note the conjugate points along the waveguide axis in (c). (b) Enlarged close up of the rays in fig (a) near the source showing the initial linear divergence. Note the caustic formation in (a) and (b) which does not occur in fig. 1, A(e). This is due to the fact that far from the waveguide axis the concavity of $w$ changes. (d) Enlarged version of (c) for easy viewing.

5. Same situation as in figure 4 with the exception that $w_0 = 10$m/s which increases the concavity. The effect of this on figure (c) is an increased number of conjugate points and fig (b) clearly illustrates initial exponential divergence.

6. Example of an environment with piecewise linear $c$ and $w$ along with a sample ray. The medium is divided horizontally, each horizontal section labeled as Region $i$, $i = 1, 2$, *etc*. Each piece of the ray which crosses two consecutive boundaries between layers is labeled segment $i$, $i = 1, 2$, *etc*.

7. Sample plot of the deviation vector, $Y$, as a function of affine parameter, $\lambda$, for a piecewise linear SSP and $w = 0$. In this case the deviation vector depends linearly on $\lambda$ along each segment of the ray. The dotted lines separate individual segments.

8. Ray fan incident on an inhomogeneous half space described in section IVc EXAMPLE. The complete ray fan in both the homogeneous and inhomogeneous regions is illustrated. The caustic, displayed in bold, was plotted from the exact solution presented in this section.

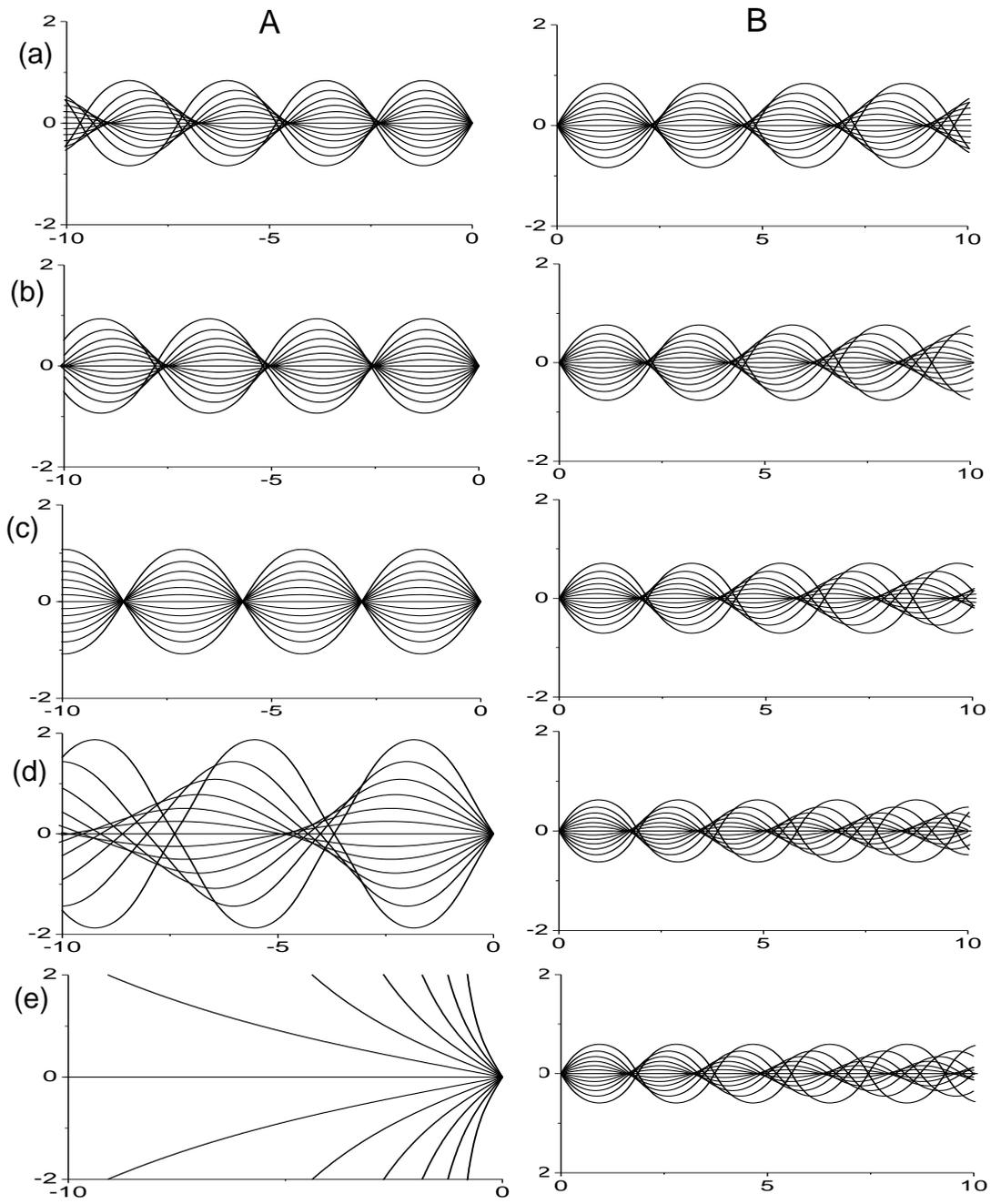

Figure 1

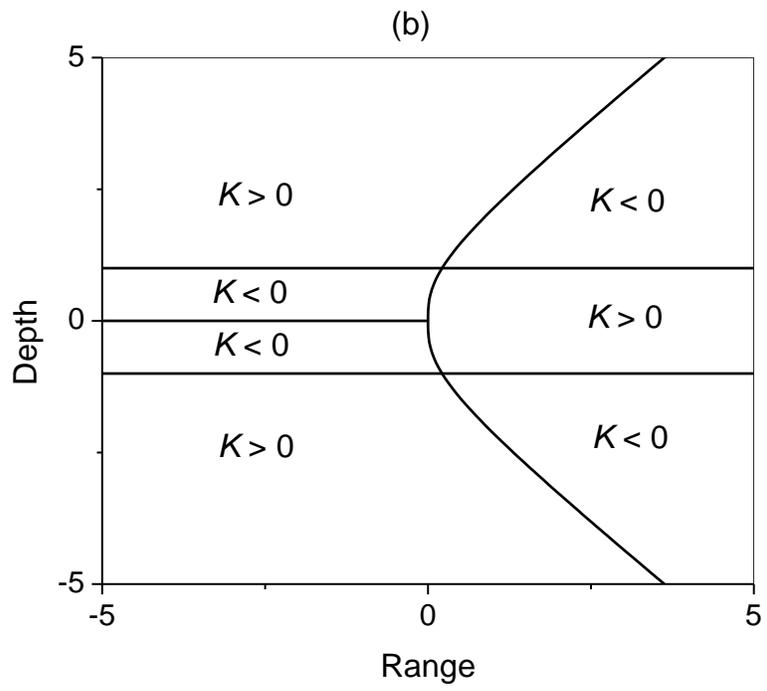

(b)

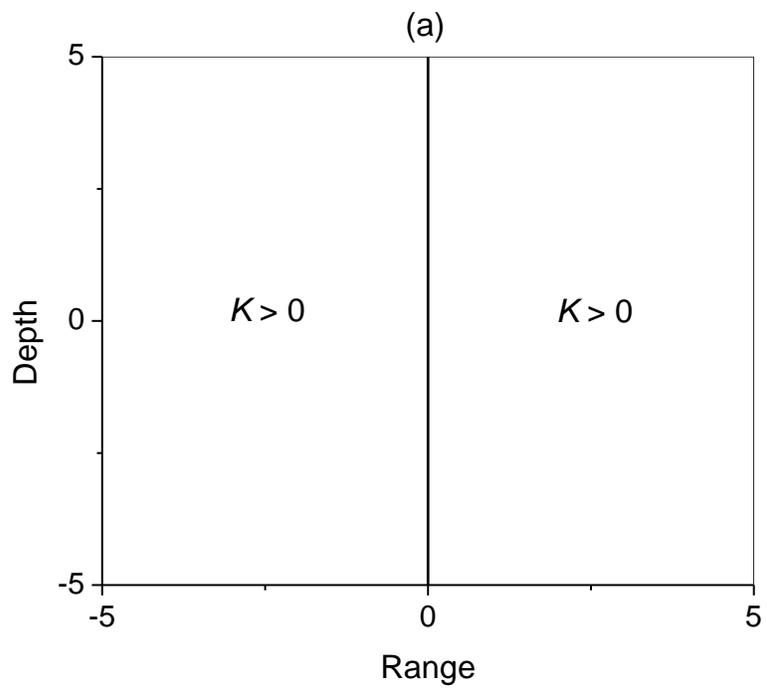

(a)

Figure 2

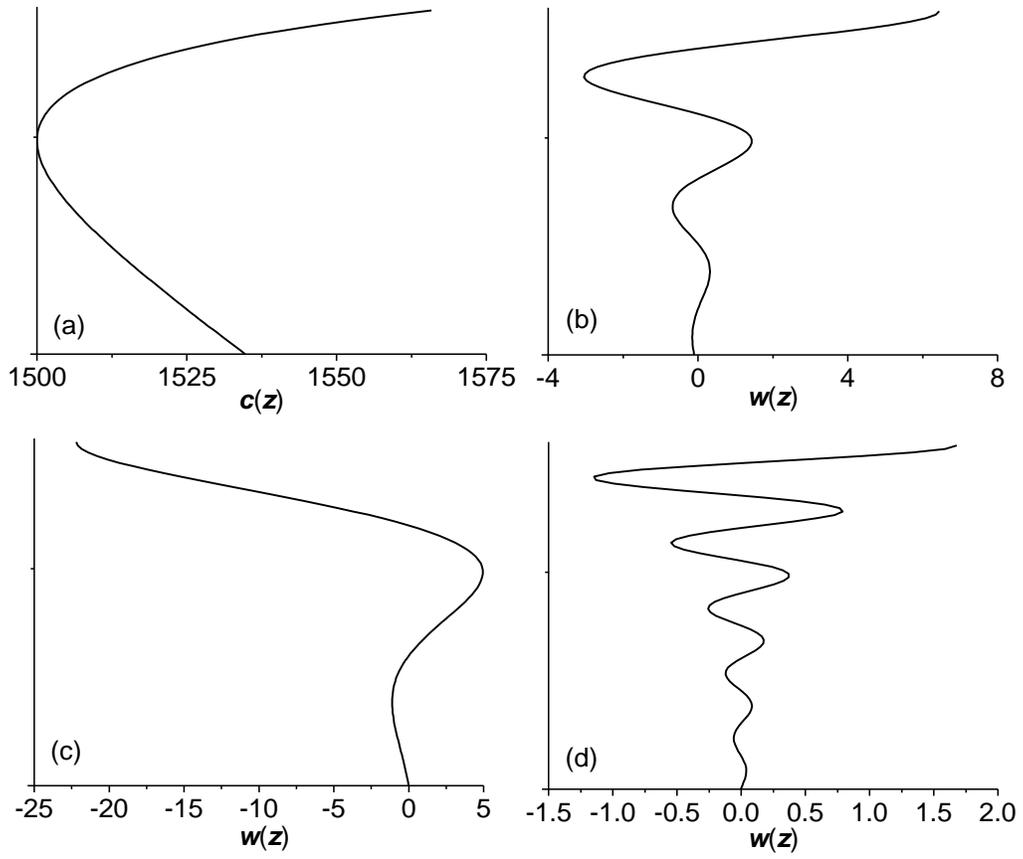

Figure 3

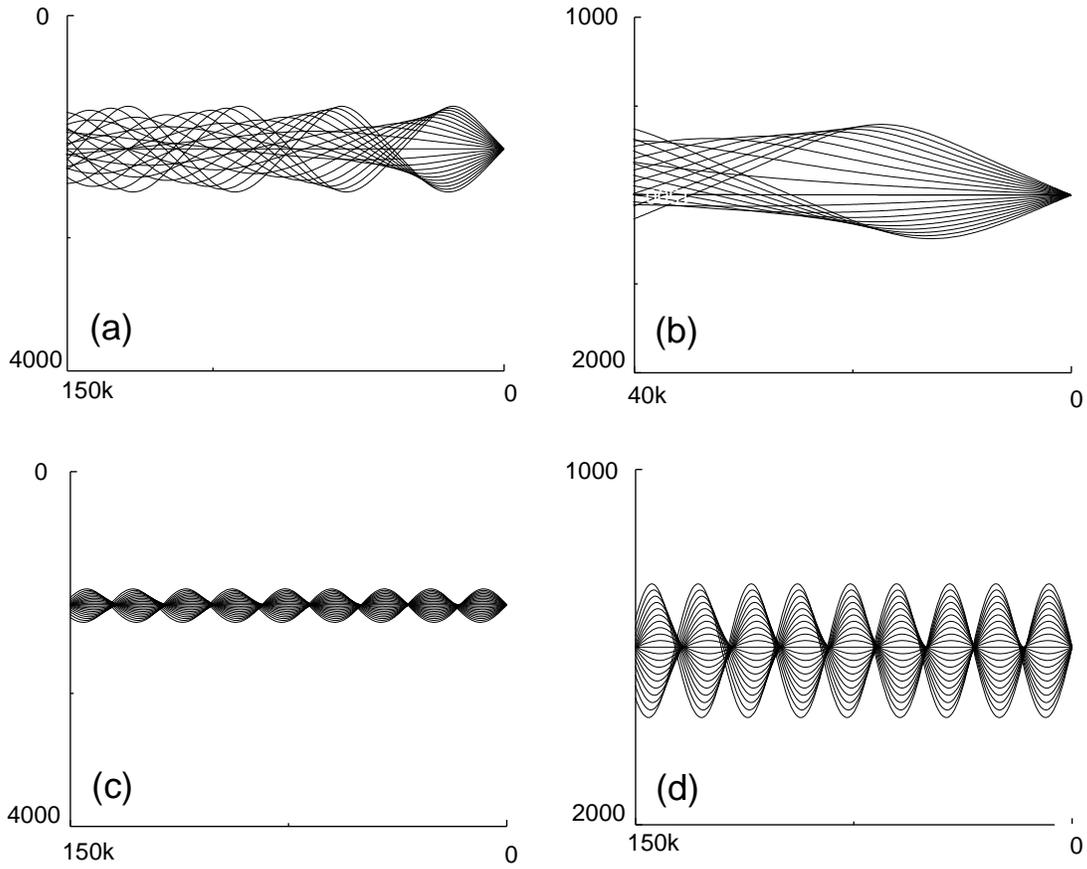

Figure 4

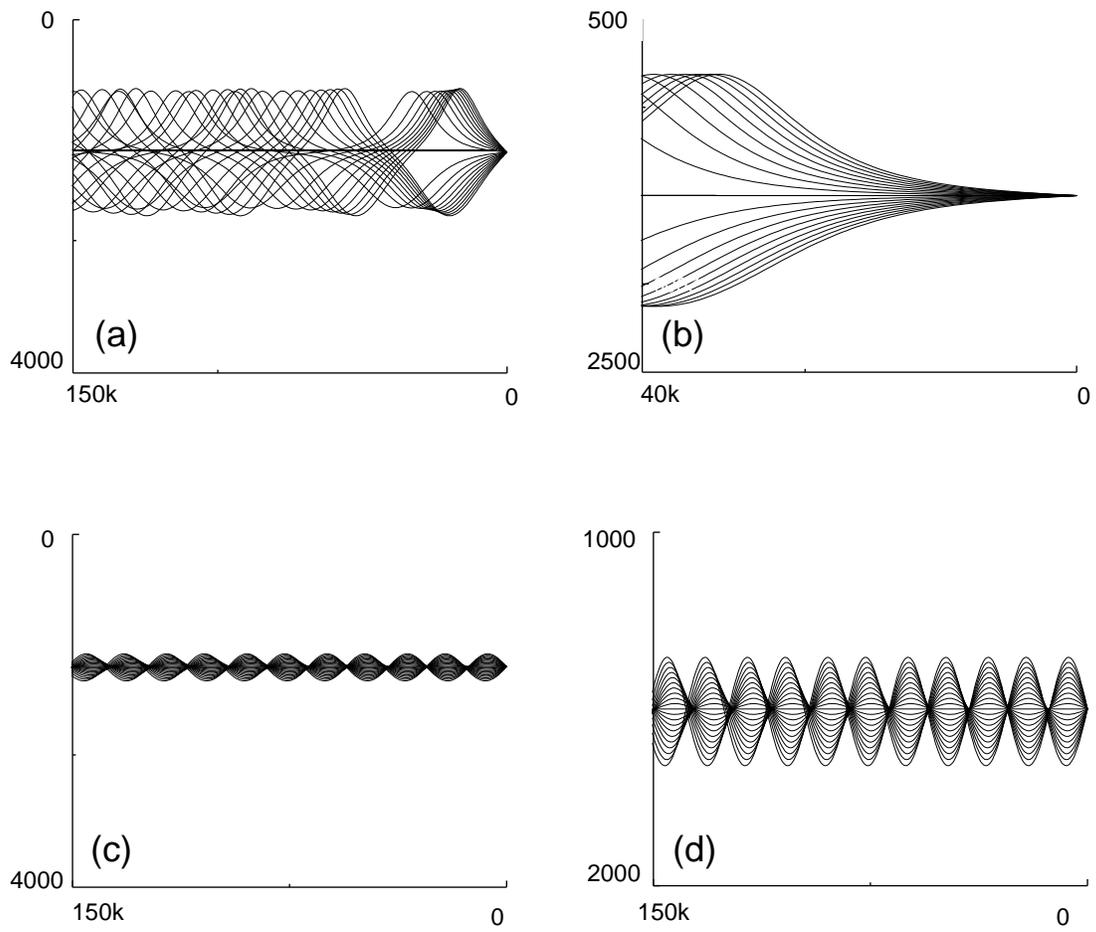

Figure 5

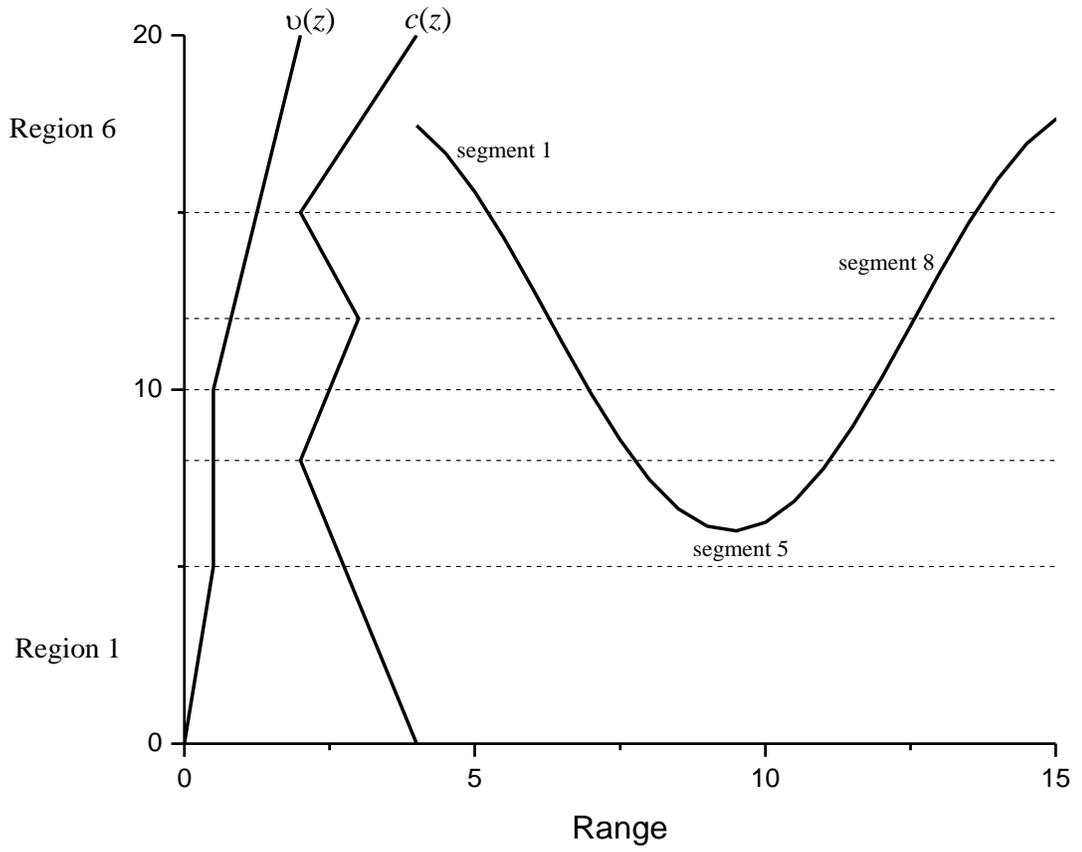

Figure 6

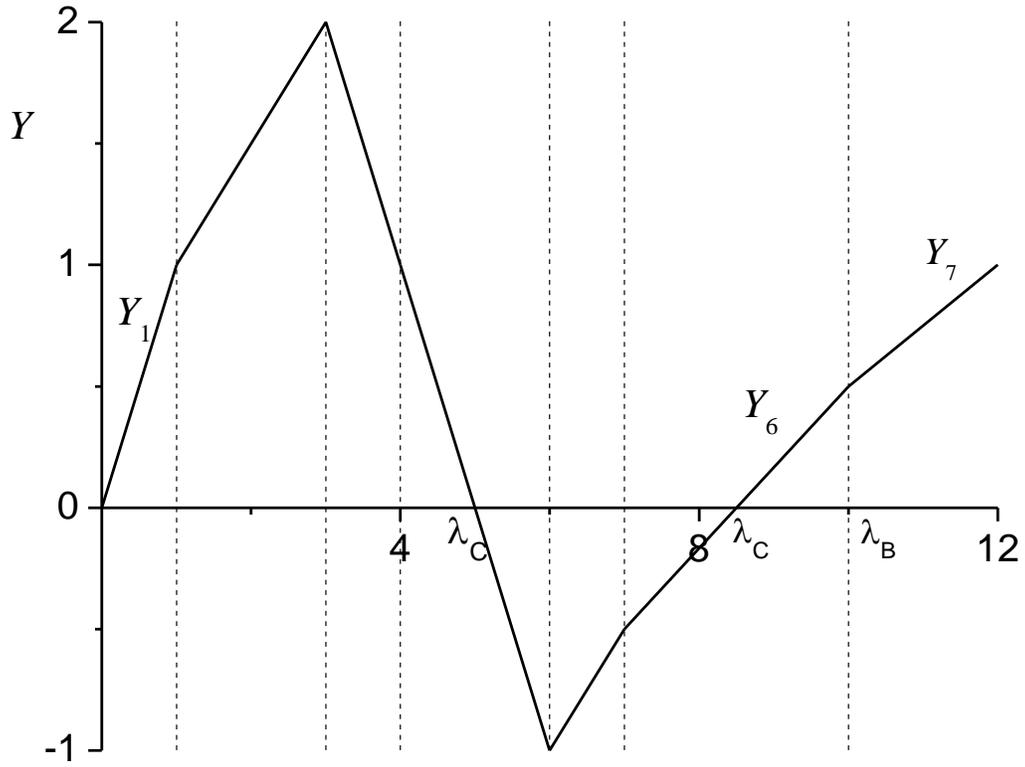

Figure 7

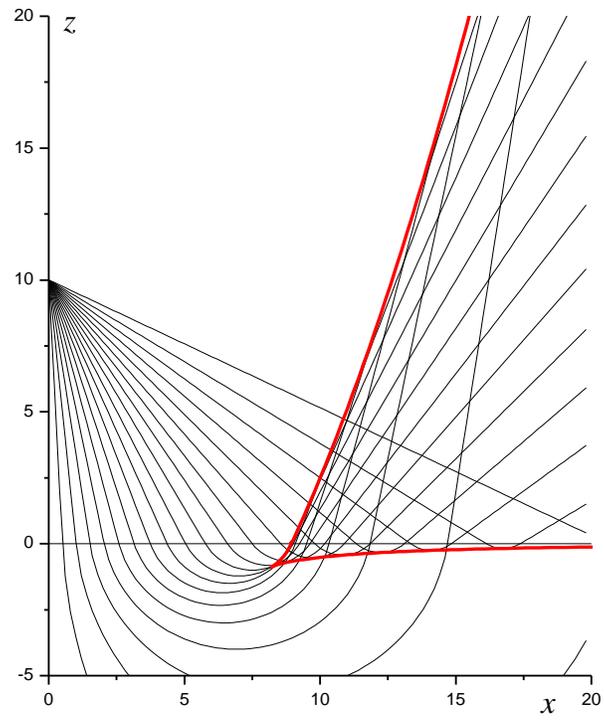

Figure 8